\documentstyle[twocolumn,aps,floats,graphicx]{revtex}

\begin{document}

\newcommand{\bec}{\begin{center}}
\newcommand{\ec}{\end{center}}
\newcommand{\be}{\begin{equation}}
\newcommand{\ee}{\end{equation}}
\newcommand{\beqn}{\begin{eqnarray}}
\newcommand{\eeqn}{\end{eqnarray}}
\newcommand{\bet}{\begin{table}}
\newcommand{\ent}{\end{table}}
\newcommand{\bib}{\bibitem}

\wideabs{

\title{
The checkerboard modulation and the inter-layer asymmetry of the hole density in cuprates
}

\author{P. S\"ule} 
  \address{Research Institute for Technical Physics and Material Science,\\
Konkoly Thege u. 29-33, Budapest, Hungary,\\
sule@mfa.kfki.hu
}

\date{\today}

\begin{abstract}

\maketitle

The 2D pair-condensate is characterized by a charge ordered state
with a "checkerboard" pattern in the planes and with an alternating superstructure along the c-axis.
We find that Coulomb energy gain occurs along the $c$-axis, which is proportional to the measured condensation energy ($U_0$) and to $T_c$:
$E_c^{3D} \approx 2 (\xi_{ab}/a_0+1)^2 U_0 \approx k_B T_c$ and is due to inter-layer charge complementarity (charge asymmetry of the boson condensate)
where $\xi_{ab}$ is the coherence length of the condensate and $a_0 \approx 3.9 \AA$ is the in-plane lattice constant.
The static $c$-axis dielectric constant $\epsilon_c$ and the coherence length $\xi_{ab}$ are also calculated for various cuprates
and compared with the available experimental data and the agreement is excellent.
 Theories and models of superconducting state\\

\end{abstract}
}

\section{Introduction}

  Charge ordering seems to be a general phenomenon in various layered
metal oxides \cite{co_oxides}.
 Among these materials charge ordered state (COS) is also found in superconducting
cuprates, although its role plays in high temperature superconductivity (HTSC) is still a
matter of considerable debate \cite{co_cuprates}.
Local probes such as scanning tunnelling microscopy (STM) revealed recently evidence for a
static checkerboard charge pattern with a real-space modulation periodicity of $4a_0$ in the
vortex core of Bi2212 \cite{Hoffman} which is a provocative evidence for pinned
charge stripes \cite{Kivelson}.
Although the precise period of the modulation is still controversial, however, it seems to be comparable
with the lattice constant \cite{Kivelson}.

  The checkerboard COS of cuprates attracted recently the attention of several theoreticians as well
\cite{Alexandrov}.
The checkerboard COS (CCOS) can be understood as the in-plane alternation of holes
and anti-holes (electron-hole pairs) in such a way that the Cooper wave-function is composed of the anti-holes and is localized
within the coherence area \cite{Sule:condmat}.
 Therefore the black and white "fields" of the CCOS correspond to partial charges of $\sim \pm 0.16e$ (pinned to lattice sites), that is the
optimal hole content found in various cuprates \cite{Presland,Sule}.
Using this simple static picture
the sum of the $\sim -0.16e$ partial charges (anti-holes, e.g. the black fields) on the checkerboard gives the $2e$ charge
of the charge carrier quasiparticle (charge sum rule for the CCOS).

 Starting from a Cu-O bond oriented "checkerboard" charge pattern with the observed $4a_0$ periodicity
\cite{Hoffman}
  we propose a simple phenomenological model for explaining the 3D character of HTSC
in cuprates supported by calculations.
Within our model the width of the checkerboard coincides with the superconducting
coherence length $\xi_{ab}$.
Furthermore we assume the alternation of the "checkerboard" charge pattern along the c-axis (that is normal
to the planes) which leads to Coulomb energy gain. Without this assumption inter-layer (IL) {\em Coulomb instabilty}
occurs in layered cuprates due to the enormous IL repulsion of holes.
We would like to study the magnitude of direct Coulomb interaction
between charge ordered square superlattice layers as a possible source of pairing interaction.
Our intention is to understand HTSC within the context of an IL Coulomb-mediated mechanism.
 The IL charging energy we wish to calculate depends on the IL spacing ($d$), the IL dielectric
constant $\epsilon_c$, the hole content $p$ and the size of the superlattice.
Finally we calculate the static $c$-axis dielectric constant $\epsilon_c$ and the coherence length $\xi_{ab}$ for various cuprates
which are compared with the experimental observations.

\section{checkerboard charge pattern and the supercondcuting order parameter}

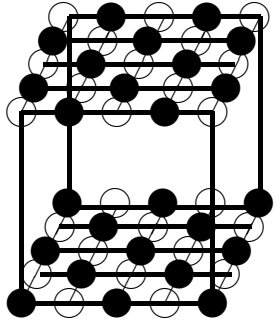
\begin{figure}

\setlength{\unitlength}{0.05in}
\begin{picture}(30,30)(-22,4)
\linethickness{0.55mm}
  \put(0,0){\line(1,2){5.0}}
  \put(20,0){\line(1,2){5.0}}
  \put(0,20){\line(1,2){5.0}}
  \put(20,20){\line(1,2){5.0}}
  \put(5,0){\line(1,2){5.0}}
  \put(10,0){\line(1,2){5.0}}
  \put(15,0){\line(1,2){5.0}}
  \put(5,20){\line(1,2){5.0}}
  \put(10,20){\line(1,2){5.0}}
  \put(15,20){\line(1,2){5.0}}

  \put(20,0){\line(0,1){20.0}}
  \put(0,0){\line(0,1){20.0}}
  \put(0,20){\line(1,0){20.0}}
  \put(0,0){\line(1,0){20.0}}
  \put(5,30){\line(1,0){20.0}}
  \put(5,10){\line(0,1){20.0}}
  \put(25,10){\line(0,1){20.0}}
  \put(5,10){\line(1,0){20.0}}
  \put(2,3){\line(1,0){20.0}}
  \put(2.5,5.5){\line(1,0){20.0}}
  \put(4,8.){\line(1,0){20.0}}
  \put(1.3,22.5){\line(1,0){20.0}}
  \put(2.3,25.0){\line(1,0){20.0}}
  \put(3.5,27.5){\line(1,0){20.0}}
  \put(0,0){\circle*{3}}
  \put(5,0){\circle{3}}
  \put(10,0){\circle*{3}}
  \put(15,0){\circle{3}}
  \put(20,0){\circle*{3}}
  \put(0,20){\circle{3}}
  \put(5,20){\circle*{3}}
  \put(10,20){\circle{3}}
  \put(15,20){\circle*{3}}
  \put(20,20){\circle{3}}
  \put(1.6,3){\circle{3}}
  \put(6.3,3){\circle*{3}}
  \put(11.5,3){\circle{3}}
  \put(16.5,3){\circle*{3}}
  \put(21.3,3){\circle{3}}
  \put(2.5,5.5){\circle*{3}}
  \put(7.5,5.5){\circle{3}}
  \put(12.5,5.5){\circle*{3}}
  \put(17.5,5.5){\circle{3}}
  \put(22.5,5.5){\circle*{3}}
  \put(4,8){\circle{3}}
  \put(8.6,8){\circle*{3}}
  \put(13.8,8){\circle{3}}
  \put(18.8,8){\circle*{3}}
  \put(23.8,8){\circle{3}}
  \put(4.8,10.5){\circle*{3}}
  \put(9.8,10.5){\circle{3}}
  \put(14.8,10.5){\circle*{3}}
  \put(19.8,10.5){\circle{3}}
  \put(24.8,10.5){\circle*{3}}
  \put(1.3,22.5){\circle*{3}}
  \put(6.3,22.5){\circle{3}}
  \put(10.8,22.5){\circle*{3}}
  \put(16.4,22.5){\circle{3}}
  \put(21.4,22.5){\circle*{3}}
  \put(2.3,25.0){\circle{3}}
  \put(7.3,25.0){\circle*{3}}
  \put(12.3,25.0){\circle{3}}
  \put(17.3,25.0){\circle*{3}}
  \put(22.2,25){\circle{3}}
  \put(3.3,27.5){\circle*{3}}
  \put(8.5,27.5){\circle{3}}
  \put(13.2,27.5){\circle*{3}}
  \put(18.3,27.5){\circle{3}}
  \put(23,27.5){\circle*{3}}
  \put(4.4,30.0){\circle{3}}
  \put(9.4,30.0){\circle*{3}}
  \put(14.4,30.0){\circle{3}}
  \put(19.4,30.0){\circle*{3}}
  \put(24.4,30.0){\circle{3}}
\end{picture}

\vspace{1cm}
\caption{\small The alternating "checkerboard" charge ordered state of the hole-anti-hole condensate
in the $4a_0 \times 4a_0$ charge ordered bilayer superlattice model.
Each lattice sites (opened and filled circles) correspond to a $CuO_2$ unit cell. 
Note the charge asymmetry between the adjacent layers. The bilayer can accomodate a pair of boson condensate ($4e$).
The inter-layer charge complementarity of the charge ordered state is crucial for getting inter-layer Coulomb
energy gain.
}

\end{figure}

  The {\em supercondcuting order parameter} (OP) which corresponds to the model with a checkerboard charge modulation in the planes takes the form of
\be
 \Psi(x,y)=n_0^{1/2} [cos(\frac{x}{a_0} \pi)+cos(\frac{y}{a_0} \pi)],
\label{OP}
\ee
where the $x$ and $y$ coordinates
are varying in the range of $x,y=[0;\xi_{ab}]$. 
For simplicity the distribution of the OP is neglected in the 3rd dimension and a nearly perfect
2D character is attributed to the condensate.
The 3D anisotropy of the condensate is negligible in the superconducting (SC) state 
which is reflected by the ratio of the in-plane and out of plane coherence lengths $\xi_{ab}/\xi_c \approx 10$ \cite{Tinkham}.
The factor $n_0$ is the maximal value of the charge density at the lattice site centers.
Eq. ~(\ref{OP}) is displayed for the coherence area in FIG 2.
The modulation of the order parameter corresponds to the real-space modulation of the
hole density in the supercondcuting (SC) state.
This kind of order parameter is given earlier by Alexandrov \cite{Alexandrov2}. 

 The order parameter must satisfy the charge sum rule for the boson condensate 
indicating the localization of the pair condensate within the coherence area,
\be
 2 \approx \int_0^{\xi_{ab}^2} \vert \Psi(x,y) \vert^2 dx dy.
\label{OPsumrule}
\ee
Another restriction on $\Psi(x,y)$ is that its integral over 
a unit cell whith the area of $\sim (a_0/2)^2$ must correspond to the hole content,
\be
 \vert q_{h(ah)} \vert=\int_0^{(a_0/2)^2} \vert \Psi(x,y) \vert^2 dx dy \approx 0.16.
\label{sumrulehole}
\ee
This equation reflects the lattice site centered localization of the holes and anti-holes
and leads to a simple electrostatic model where the charge modulation for simplicity is replaced
by point charges centered in the center of the holes and anti holes (see later).

  IL Coulomb energy gain occurs only in that case when 
holes in one of the layers are in proximity with anti-holes in the other layer
(FIG 1, IL {\em electrostatic complementarity}, bilayer $5 \times 5$ ($4a_0 \times 4a_0$ model).
An important feature is then that
the boson condensate can be described by an IL {\em charge asymmetry}.
Therefore we assume an alternating charge pattern along the c-axis.
The IL coupling of the boson-boson pairs in the bilayer $5 \times 5$ model naturally suggests the
effective mass of charge carriers $m^{*} \approx 4 m_e$, as it was found by measurements \cite{krusin}.


\begin{figure}[hbtp]
\begin{center}
\includegraphics*[height=5.5cm,width=7.5cm]{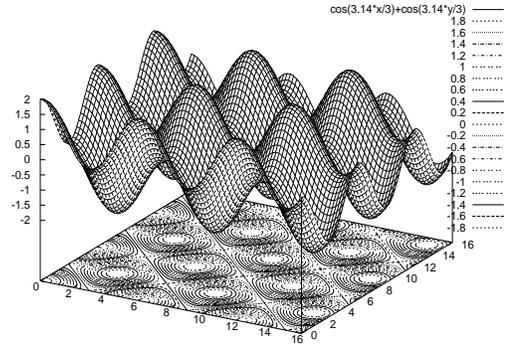}
\caption[]{
The 3D view of
the "checkerboard" charge pattern of the boson condensate
corresponding to the $4a_0$ charge modulation represented
by the order parameter given in Eq. (1).
$x$ and $y$ coordinates are given in $\AA$.
The wells and peaks correspond to holes and anti-holes, respectively.
}  
\label{cond_tc}
\end{center}
\end{figure}


The $5 \times 5$ model can be generalized to represent a $N \times N$ coherence
area where $N$ is the real space periodicity of the superlattice.
In the $N \times N$ superlattice model the hole and anti-hole partial charges at each $CuO_2$ lattice
sites are $q=\pm 4e/N^2$. Important to note that the charge sume rule
holds for the characteristic bilayer with a coherence area $\sum_i^{N^2} q_i^{ahole}=4e$
where $q_i^{ahole}$ is the partial anti-hole charge at the $i$th anti-hole lattice site.
In other words a pair of a boson condensate can be localized within a characteristic bilayer depicted in FIG 1.

 The alternation of the charge pattern along the c-axis is reflected in the order parameter
by the layer by layer alternation of $cos$ and $sin$ functions. For example, the adjacent layer
is described by

 Without any experimental evidences we assume in this paper that the IL Coulomb energy gain  
is the main contribution to the condensation energy of the supercondcuting state.
This seems to be rather arbitrary assumption, however, we see no clear cut evidence for
the $ab$-plane contribution to the condensation energy $U_0$ (given per $CuO_2$ unit cell) as well.
In this paper we study the case when HTSC is governed purely by the $c$-axis Coulomb energy
gain (potential
energy driven
superconductivity) and therefore there is no relation to the kinetic energy driven 
mechanism proposed by several authors \cite{PWA}.

 The condensation energy of a bilayer with a coherence area in the planes ($U_{0b}$) is given as follows
\be
U_{0b}=2 (n+1) \biggm[ \frac{\xi_{ab}}{a_0}+1 \biggm]^2 U_0 \approx E_c^{IL,SC},
\label{gain_sc}
\ee
where $E_c^{IL,SC}$ is the Coulomb energy gain in the SC state.
$U_0$ is the experimental condensation energy given per unit cell. 
Eq.~(\ref{gain_sc}) is generalized for multilayer cuprates introducing
$n$.
For single layer cuprates $n=0$, for bilayers $n=1$, etc.
The factor $\biggm[ \frac{\xi_{ab}}{a_0}+1 \biggm]^2$ is the number of $CuO_2$ lattice sites
in the planes (within the coherence area).
Factor $2$ is applied in Eq.~(\ref{gain_sc}) because we calculate the condensation
energy of a bilayer.

 The IL Coulomb energy is in the SC state
\be
E_c^{IL,SC}=\frac{e^2 Q}{4 \pi \epsilon_0 \epsilon_c},
\label{IL}
\ee
where
\be
Q= \sum_{m=2}^{N_l} \sum
_{ij}^{N^2} \frac{q_i^{(n)} q_j^{(m)}}{r_{ij}^{(n,m)}},
\ee
where $r_{ij}^{(n,m)}$ is the inter-point charge distance and $r_{ij}^{(n,m)} \ge d_{IL}$, where $d_{IL}$ is
 the IL distance ($CuO_2$ plane to
plane, $i \ne j$). $n,m$ represent  
layer indexes ($n \ne m$).
$q_i^{(n)}$ and  $q_j^{(m)}$ are the point charges of holes and anti-holes centered at $CuO_2$ lattice
sites in the $n$th and $m$th layers.
First the summation goes within the bilayer up to $N^2$ (the real space periodicity of the CCOS, $N \approx \frac{\xi_{ab}}{a_0}+1$)  then the IL Coulomb interaction of the
basal bilayer are calculated with other layers
along the $c$-axis in both direction ($n=1,2$).
$N_l$ is the number of layers along the $c$-axis. When $N_l \rightarrow \infty$, bulk $E_c^{IL,SC}$ is calculated.

\begin{table}
\caption[]
{The calculated coherence length of the pair condensate
given in $a_0$ using the experimental condensation energies
of various cuprates and Eq.~(\ref{N}) at optimal doping.
}
{\scriptsize
\begin{tabular}{cccccc}
 & $T_c$ (K) & $k_B T_c$ (meV) & $U_0$ ($\mu eV/u.c.$)  & $\xi_{ab}^{calc} (a_0)$ & $\xi_{ab}^{exp} (a_0)$ \\ 
\hline
  Bi2201 & 20 & 1.6 & $10^a$  &  $\sim 8$ &       \\
 LSCO   & 39  & 2.5 & $21^b$ &  $\sim 7$ &  $5-8^c$  
 \\
 Tl2201 & 85 & 7 & $100 \pm 20^d$ & $\sim 5$ & \\
 Hg1201 & 95 & 7.8 & $80-107^e$ &  $\sim 5$ & $5^f$ \\
  YBCO  & 92 & 7.5 & $110^g$ &  $\sim 3$ & $ 3-4^h $ \\
  Bi2212 & 89 & 7.3 & $95^g$ &  $\sim 3-4$ & $4-6^i$\\
\end{tabular}}
{\scriptsize
$a_0 \approx 3.88 \AA$,
$U_0$ is the measured condensation energy of various cuprates
in $\mu$ eV per unit cell at optimal doping.
$^a$ from \cite{MarelPC},
$^b$ $U_0 \approx 2$ J/mol from \cite{Loram,Momono},
$^c$ from \cite{Tinkham},
$^d$ \cite{Tsvetkov},
$^e$ $U_0 \approx 12-16$ mJ/g from \cite{Billon,Kirtley} and $\xi_{ab}$ from \cite{Thompson},
$^d$ $U_0 \approx 11$ J/mol from \cite{TallonLoram},
$^e$ $U_0 \approx 10$ J/mol from \cite{TallonLoram},
$^f$ from \cite{Thompson},
$^g$ from \cite{TallonLoram},
$^i$ from recent measurements of Wang {\em et al.}, $\xi_{ab} \approx 23 \AA (\sim 5-6 a_0)$ \cite{Wang_sci},from STM images of ref. \cite{Hoffman} $\xi_{ab}\approx 4a_0$,
$^h$ from \cite{Tinkham},
$\xi_{ab}^{calc}$ is calculated according to Eq.~(\ref{N}) and is also given in Table ~\ref{tab1} and $\xi_{ab}^{exp}$ is
the measured in-plane coherence length given in $a_0 \approx 3.9 \AA$.
The notations are as follows for the compounds:
Bi2201 is $Bi_2Sr_2CuO_{6+\delta}$,
LSCO ($La_{1.85}Sr_{0.15}CuO_4$),  
Tl2201 ($Tl_2Ba_2CuO_6$), 
Hg1201 ($HgBa_2CuO_{4+\delta}$),  
 YBCO ($YBa_2CuO_7$) and
 Bi2212 is $Bi_2Sr_2CaCu_2O_{8+\delta}$.
}
\label{tab1}
\end{table}

\section{The condensation energy and $T_c$}

  The plot of $U_{0b}$ (the bilayer condensation energy) against $T_c$ is shown
in FIG 3 using only experimental data.
Remarkably the data points of various cuprates with a variety of critical temperature
fit to a line and its slope $U_{0b}/T_c$ is the Boltzmann constant $k_B$.

\begin{figure}[hbtp]
\begin{center}
\includegraphics*[height=4.5cm,width=6.5cm]{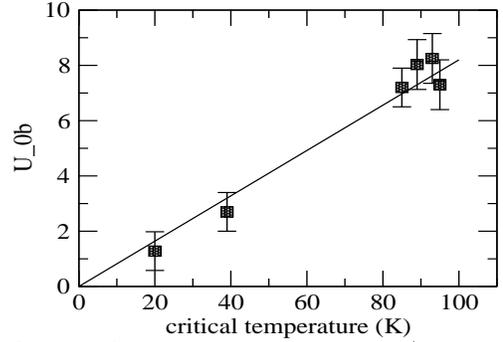}
\caption[]{
The bilayer condensation energy ($U_{0b}$, meV)
as a function of
the critical temperature (K) at optimal doping. 
The straight line is a linear fit to the data. The slope of the linear fit
is $U_{0b}/T_c \approx k_B$ which is a strong evidence of Eq.~(\ref{kbtc}).
The error bars denote standard deviations estimated from various measurements of
the condenstaion energy.
}
\label{tc_cond}
\end{center}
\end{figure}

The average value we get is $k_B \approx 1.3 \pm 0.2 \times 10^{-23} J/K$ which is
remarkably close to the value of $k_B=1.38 \times 10^{-23} J/K$.
In the rest of the paper we will present further evidences in order to show
that the agreement might not be accidental.
According to the correlation found between $U_{0b}$ and $T_c$ 
the following formula can be derived
using Eq.~(\ref{gain_sc}),

\be
U_{0b}=2 (n+1) \biggm[ \frac{\xi_{ab}}{a_0}+1 \biggm]^2 U_0 \approx k_B T_c.
\label{kbtc}
\ee
Therefore the bilayer condensation energy $U_{0b}$ can directly be related
to the thermal motion at $T_c$.
Although the mechanism of thermally induced depairing is not understood yet,
it might be due to the destruction of the lattice-CCOS interactions above $T_c$.
In this respect polarons or bipolarons might play a role in HTSC \cite{Alexandrov}.

 In order to test the validity of Eq.~(\ref{kbtc}) we estimate the 
coherence length of the pair condensate derived from Eq.~(\ref{kbtc})
and using only experimental data,
\be
\xi_{ab} \approx a_0 \biggm[ \sqrt{\frac{k_B T_c}{2 (n+1) U_0}}-1 \biggm]
\label{N}
\ee
The results are given in Table~\ref{tab1} as $\xi_{ab}^{calc}$ (in $a_0$ unit) and compared with the available
measured $\xi_{ab}^{exp}$. The agreement is excellent which strongly suggests that Eq.~(\ref{N})
should also work for other cuprates.

 The expression Eq.~(\ref{kbtc}) leads to the very simple formula for the critical temperature
using Eq.~(\ref{gain_sc}) and a simple Coulomb expression for the IL coupling energy $E_c^{IL,SC}$
($T_c \approx k_B^{-1} E_c^{IL,SC}$)
\be
T_c (N,d,\epsilon_c) \approx \frac{e^2 Q}{4 \pi \epsilon_0 \epsilon_c k_B}  
\label{kbtc2}
\ee
When $N_l \rightarrow \infty$, bulk $T_c$ is calculated.
$T_c$ can also be calculated for thin films when $N_l$ is finite.
$\epsilon_c$ can also be derived
\be
\epsilon_c \approx \frac{e^2 Q}{4 \pi \epsilon_0 k_B T_c}  
\label{epsc}
\ee
where a $c$-axis average of $\epsilon_c$ is computed when $N_l \rightarrow \infty$.

\begin{table}
\caption[]
{The calculated dielectric constant $\epsilon_c$ using Eq.~(\ref{epsc})
in various cuprates at the calculated coherence length $\xi_{ab}$ of the charge ordered state
given in Table I.
}
{\scriptsize
\begin{tabular}{lccccc}
 & $d (\AA)$ & $T_c (K)$ & $\xi_{ab}^{calc} (a_0)$ & $\epsilon_c$ & $\epsilon_c^{exp}$  \\ \hline
 Bi2201 &    12.2 & 20   &   8  &  9.9 & 12$^a$ \\
 LSCO &      6.65 &  39 & 7 &  11.3 & $23 \pm 3, 13.5^b$  \\
 Hg1201    &     9.5   & 95  & 5 & 6.5 &    \\
 Tl2201 &    11.6 &  85 &  5 & 13.0 & 11.3$^c$ \\
  YBCO  &   8.5  & 93  &  3 & 19.4 & 23.6$^d$ \\ 
\end{tabular}}
{\scriptsize
where $\xi_{ab}^{calc}$ is the estimated in-plane coherence length given in $a_0 \approx 3.9 \AA$ unit.
$d$ is the $CuO_2$ plane to plane inter-layer distance in $\AA$, $T_c$ is the experime
ntal critical temperature.
$\epsilon_c$ is from Eq.~(\ref{epsc}).
$\epsilon_c^{exp}$ are the measured values obtained from the following references: 
$^a$ \cite{Boris}, 
$^b$ \cite{epsilonc}, or from reflectivity measurements \cite{Tsvetkov},
$\omega_p \approx 55 cm^{-1}$ \cite{Sarma}, $\lambda_c \approx 3 \mu$m \cite{Kirtley},
$^c$ from \cite{Tsvetkov}, 
$^d$ from reflectivity measurements: $\omega_p \approx 60 cm^{-1}$ \cite{Sarma}, $\lambda_c \approx 0.9 \mu$m \cite{Kitano2}.
}
\label{tab2}
\end{table}

The calculation of the $c$-axis dielectric constants $\epsilon_c$ might provide
further evidences for Eq.~(\ref{kbtc}) when compared with the measured values \cite{Kitano,epsilonc}.
In Table ~\ref{tab2} we have calculated the static dielectric function $\epsilon_c$ using
Eq.~(\ref{epsc}) and compared with the experimental impedance measurements \cite{Cao,epsilonc}.
$\epsilon_c$ can also be extracted from the $c$-axis optical measurements using the
relation \cite{Tsvetkov}
$\epsilon_c(\omega) =\epsilon_c(\infty)-c^2/(\omega_p^2 \lambda_c^2)$, 
where $\epsilon_{\infty}$ and $\omega_p$ are the high-frequency dielectric constant and the plasma frequency, respectively \cite{Tamasaku}. $c$ and $\lambda_c$ are the speed of light
and the $c$-axis penetration depth. 
At zero crossing $\epsilon_c(\omega)=0$ and $\omega_p=c/(\lambda_c \epsilon_c^{1/2}(\infty))$.
Using this relation we predict for the single layer Hg1201 the low plasma frequency of
$\omega_p \approx 8$ cm$^{-1}$ using $\epsilon_c=\epsilon_c(\infty) \approx 6.5$ (Table II) and $\lambda_c \approx
8$ $\mu$m \cite{Kirtley}.

\section{Conclusion}

  Our primary result is that the boson condensate in the superconducting state can be described
by an alternating "checkerboard" type of charge pattern along the c-axis which leads to 
inter-layer charge complementarity and to Coulomb energy gain. 
Our proposal is that this gain is converted to the condensation energy of the superconducting state, although the detailed mechanism of this process
still remains unclear. 
Within our model the pairing glue is provided by inter-layer coupling.
This phsyical picture naturally explains the variation of $T_c$ system by system in various
conditions (external and chemical pressure, multilayers, heterostructures etc.).
Although the alternation of the charge pattern along the c-axis is not yet seen experimentally
the model might be useful for further studies in the future.
The reason for this is simple: without the assumption of inter-layer charge complementarity the superconducting state
would "suffer" from enormous inter-layer Coulomb instability (repulsion) which is certainly not the case.
Local probes with sufficient depth resolution in thin films might detect the presence of
such supermodulation of the charge pattern along the $c$-axis in the future if the phenomenon
exists in nature.
  Unfortunatelly the microscopic behaviour of pairing is still puzzling: the possible conversion of the
$c$-axis free energy gain to pairing needs further understanding. 
Anyhow the modulation of the charge density along the $c$-axis theoretically provides the possibilty of better understanding
HTSC.


{\small
It is a privilige to thank M. Menyh\'ard for the continous support.
I greatly indebted to E. Sherman for reading the manuscript carefully
and for the helpful informations.
I would also like to thank for the helpful discussions with T. G. Kov\'acs,
I. Bozovic and for the stimulating comments to A. S. Alexandrov. 
This work is supported by the OTKA grant F037710
from the Hungarian Academy of Sciences}
\\


\end{document}